\documentclass[preprint,review,12pt]{elsarticle}

\usepackage{lineno,hyperref}
\modulolinenumbers[5]

\journal{Physica B: Condensed Matter}

\usepackage{graphicx}% Include figure files
\usepackage{color}

\usepackage{amssymb}
\usepackage{amsmath}

\begin{document}
	
\begin{frontmatter}

\title{Temperature-dependent local structure and lattice dynamics of 1T-TiSe$_2$ and 1T-VSe$_2$ probed by X-ray absorption spectroscopy}

\author[ISSP]{Inga Pudza}
\ead{inga.pudza@cfi.lu.lv}

\author[ISSP]{Boris Polyakov}
\ead{boris.polyakov@cfi.lu.lv}

\author[ISSP]{Kaspars Pudzs}
\ead{kaspars.pudzs@cfi.lu.lv}

\author[DESY]{Edmund Welter}
\ead{edmund.welter@desy.de}

\author[ISSP]{Alexei Kuzmin\corref{ak}}
\ead{a.kuzmin@cfi.lu.lv}
\cortext[ak]{Corresponding author}

\address[ISSP]{Institute of Solid State Physics, University of Latvia, Kengaraga street 8, LV-1063 Riga, Latvia}

\address[DESY]{Deutsches Elektronen-Synchrotron DESY, Notkestr. 85, 22607 Hamburg, Germany}

\begin{abstract}

The local atomic structure and lattice dynamics of two isostructural layered transition metal dichalcogenides (TMDs), 1T-TiSe$_2$ and 1T-VSe$_2$, were studied using temperature-dependent X-ray absorption spectroscopy at the Ti, V, and Se K-edges. Analysis of the extended X-ray absorption fine structure (EXAFS) spectra, employing reverse Monte Carlo (RMC) simulations, enabled tracking the temperature evolution of the local environment in the range of 10-300~K. The atomic coordinates derived from the final atomic configurations were used to calculate the partial radial distribution functions (RDFs) and the mean-square relative displacement (MSRD) factors for the first ten coordination shells around the absorbing atoms.  Characteristic Einstein frequencies and effective force constants were determined for 
Ti--Se, Ti--Ti, V--Se, V--V, and Se--Se  atom pairs from the temperature dependencies of MSRDs. 
The obtained results reveal differences in the temperature evolution of lattice dynamics and the strengths of intralayer and interlayer interactions in TiSe$_2$ and VSe$_2$. 
\end{abstract}


\end{frontmatter}

%\linenumbers

\newpage

\section{Introduction}

Transition metal dichalcogenides (TMDs) have garnered significant attention in the scientific community due to their layered structure, resulting in remarkable electronic, optical, mechanical, tribological, catalytic, and magnetic properties as well as numerous possible technological applications  \cite{Bhimanapati2015, Zhan2019, Shanmugam2022, Lin2023}. Among all, the charge density wave (CDW) behavior observed for a number of TMDs MX$_2$ (M = Ti, V, Ta; X= S, Se) is particularly fascinating \cite{Ritschel2015,Hossain2017,Jolie2019}. Indeed, TMDs represent the first layered materials in which the presence of CDWs was detected \cite{Wilson1974}. The CDW state involves the periodic modulation of the electron density within the material coupled with lattice distortion, leading to changes in the material electronic and optical properties \cite{Leroux2018, Guo2021}. Models describing CDW mechanisms are based solely on crystal structure distortion/superlattice formation, exciton–phonon interactions, or electron–phonon interactions \cite{Rossnagel2011,Adam2022}.
Due to the minute nature of distortion or atomic shifts in the CDW state ($\sim$0.1-0.15~\AA), the resulting superstructure might generate only relatively faint secondary peaks within the X-ray diffraction patterns \cite{Guo2021}.

In this paper, two isostructural TMDs, namely  titanium selenide (TiSe$_2$)  and vanadium selenides (VSe$_2$), were selected to evaluate the effect of the CDW ordering on their local environments. The choice of TMDs was dictated by the feasibility of measuring high-quality X-ray absorption spectra at both the metal and chalcogenide K absorption edges.

\begin{figure}[t]
	\centering	
	\includegraphics[width=0.7\textwidth]{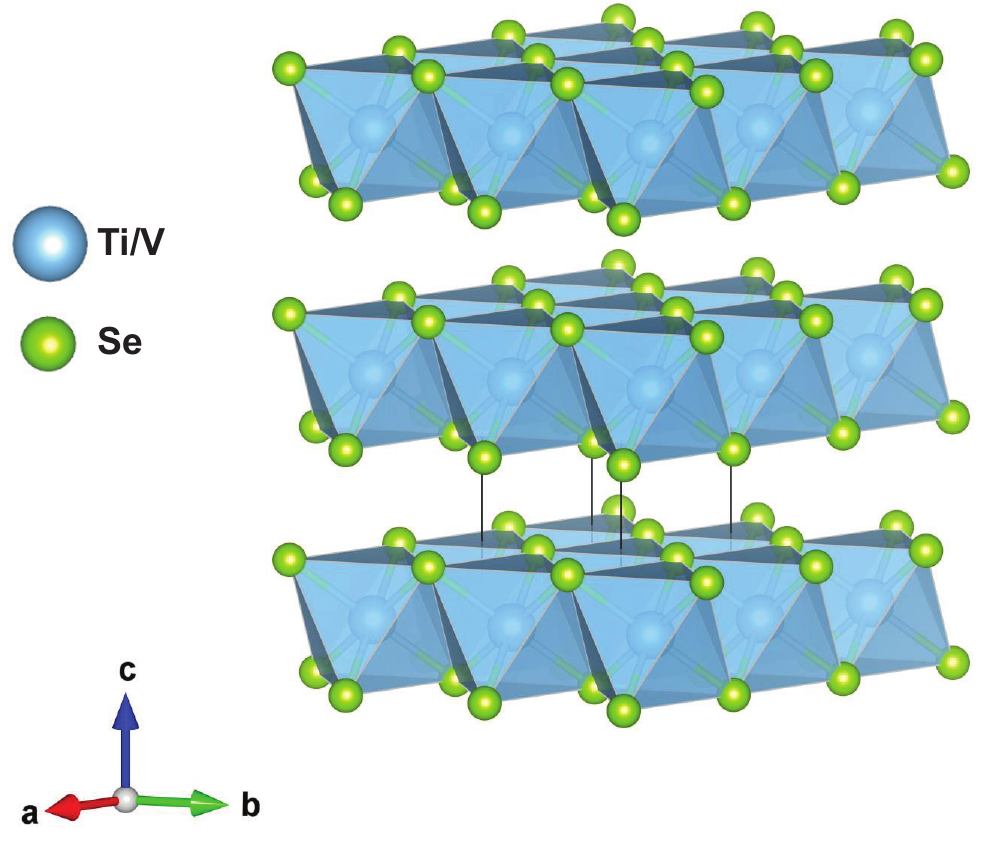}
	\caption{Crystallographic structure of trigonal (space group $P\bar{3}m1$ (164)) TiSe$_2$ (VSe$_2$) \protect\cite{Riekel1976, Hoschek1939}.}
	\label{fig1}
\end{figure}

Under ambient conditions, both TiSe$_2$ and VSe$_2$ crystallize in the 1T polytype with a space group of $P\bar{3}m1$ (164) (Fig.\ \ref{fig1}) \cite{Riekel1976, Hoschek1939}. In this structure, the Ti(V) atoms are octahedrally coordinated to six covalently bonded Se atoms, and [Ti(V)Se$_6$] octahedra are connected by edges into layers, joined by weak van der Waals (vdW) interactions.   
The interactions between neighbouring layers play an important role in assembling and exfoliation of vdW (hetero)structures and, thus, allow engineering of their electronic, optical, and mechanical properties  \cite{Liu2016,Shi2018,Bian2021}.  
While the importance of weak interlayer coupling is recognized, its accurate experimental quantification remains challenging  \cite{Shi2018,Pudza2023}.

A commensurate CDW phase in bulk 1T-TiSe$_2$ forms below $T_{\rm CDW} \approx $202~K with a 2$a \times $2$a \times $2$c$  
superlattice  \cite{DiSalvo1976,Rossnagel2011}. 
The neutron diffraction study proposed the low-temperature (at 77~K) lattice distortion due to displacements of about 0.085~\AA\ for Ti and 0.028~\AA\ for Se parallel to the plane of the layer \cite{DiSalvo1976}. 
However, more recent synchrotron X-ray scattering experiments pointed to a zone-boundary phonon 
softening mechanism of CDW in 1T-TiSe$_2$ \cite{Holt2001,Weber2011}. 

1T-VSe$_2$ undergoes an incommensurate CDW transition at $T_{\rm CDW} \approx $110~K and a commensurate CDW transition at $\approx$80~K, forming a 4$a \times $4$a \times $3$c$ superlattice \cite{Bayard1976,Eaglesham1986}. 
The origin of the CDW in VSe$_2$ was attributed to the electron-phonon interaction, based on inelastic X-ray scattering and first-principles calculations \cite{Diego2021}. Besides, an important role of  vdW forces in the CDW melting was proposed, likely caused by the  out-of-plane nature of the CDW, which modulates interlayer distance \cite{Diego2021}.
Recently, a pressure-induced CDW state was observed in VSe$_2$ at room temperature within the pressure range of 10-15~GPa
by Raman spectroscopy \cite{Feng2020} and X-ray diffraction (XRD) \cite{Guo2021}. The Se K-edge X-ray absorption spectroscopy (XAS) was also used in \cite{Guo2021} to probe local distortions but only qualitative analysis was reported.  

In this study, we employed XAS combined with the advanced data analysis methodology of extended X-ray absorption fine structure (EXAFS), based on reverse Monte Carlo (RMC) simulations coupled with an evolutionary algorithm (EA) approach \cite{Timoshenko2012rmc,Timoshenko2014rmc}, to investigate the local atomic structure and lattice dynamics in bulk 1T-TiSe$_2$ and 1T-VSe$_2$. 
We have demonstrated recently \cite{Pudza2023} that employing this approach for 2H$_c$-MoS$_2$ enables the retrieval of structural data from distant coordination shells and, thus, provides valuable information on both interlayer and intralayer coupling. Here we took advantage of high-quality EXAFS spectra measured at two (Ti/V and Se) K absorption edges and performed their simultaneous analysis to reconstruct the temperature evolution of the local environment in the corresponding TMDs in the range of 10-300~K. 
This allowed us to determine the amplitudes of thermal vibrations for atoms located in the first ten coordination shells around absorbing atoms, which were further used to evaluate the strengths of interlayer and intralayer interactions. 
We showed that despite the isostructural nature of TiSe$_2$ and VSe$_2$, the temperature-dependent evolution of their lattice dynamics has some peculiarities.

\section{Materials and methods}

\subsection{Synthesis procedure}

TiSe$_2$ and VSe$_2$ were prepared using the chemical vapor transport technique with an I$_2$ transport agent \cite{UBALDINI2014878,Pandey2020}. The starting materials titanium (or vanadium) and selenium in the form of powder were weighed out with a molar ratio of M:Se = 1:2 (M = Ti or V) and loaded into a quartz ampoule together with the iodine (molar ratio I$_2$:M = 0.05). The filled ampoule was evacuated to pressure 10$^{-5}$~torr and sealed at a length of approximately 12-14~cm using the oxygen-methane flame. For TiSe$_2$ synthesis the ampoule was heated in a two-zone furnace with $T_{\rm hot} = 700$~$^\circ$C at the hot zone and $T_{\rm cold} = 600$~$^\circ$C at the cold zone for 24~h, then naturally cooled down (for VSe$_2$ $T_{\rm hot} = 820$~$^\circ$C, $T_{\rm coldt} = 650$~$^\circ$C, heating time 72~h).

\subsection{X-ray diffraction}

\begin{figure}[t]
	\centering
	\includegraphics[width=0.5\textwidth]{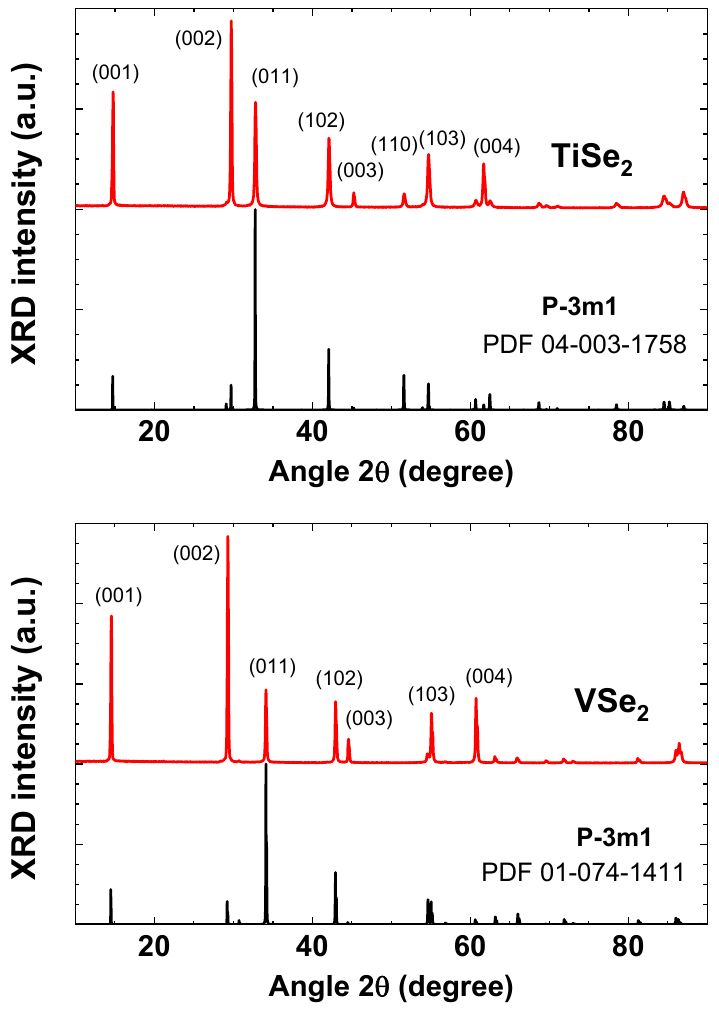}
	\caption{ Powder X-ray diffraction patterns of synthesized 1T-TiSe$_2$  and 1T-VSe$_2$. Main Bragg peaks are indexed to the $P\bar{3}m1$ (164) space group. Reference patterns corresponding to TiSe$_2$ (PDF Card 04-003-1758) and VSe$_2$ (PDF Card 01-074-1411) phases are shown for comparison. }
	\label{fig2}
\end{figure}

The phase purity of the TMD samples was confirmed using X-ray powder diffraction (XRD). 
Diffraction patterns were collected at room temperature using a benchtop Rigaku MiniFlex 600 diffractometer with Bragg-Brentano $\theta$-2$\theta$ geometry. An X-ray tube with copper anode (Cu K$\alpha$ radiation, $\lambda$ = 1.5418~\AA), operated at $U$ = 40~kV and $I$ = 15~mA, was used as a source. 
The obtained X-ray diffraction patterns of TiSe$_2$ and VSe$_2$ are compared in Fig.\ \ref{fig2} with the reference ones. All Bragg peaks were indexed to pure TiSe$_2$ (PDF Card 04-003-1758) and VSe$_2$ (PDF Card 01-074-1411) phases possessing $P\bar{3}m1$ (164) space group. Slight variations in peak intensities are noticeable, likely attributed to the favored alignment of layered structures that could arise during the process of sample preparation.
No impurity phases were detected, demonstrating the single-phase composition of the synthesized samples.

\subsection{X-ray absorption experiments}

The temperature-dependent (10-300~K) X-ray absorption spectra of bulk 1T-TiSe$_2$ and 1T-VSe$_2$ were recorded at  Ti (4966~eV), V (5465~eV) and Se (12658~eV) K-edges in transmission mode at the DESY PETRA III P65 Applied XAFS undulator beamline \cite{P65}. The storage ring operated in top-up 480 bunch mode at the energy $E$ = 6.08~GeV and current $I$ = 100~mA. The synchrotron radiation was monochromatized using a Si(111) double-crystal monochromator, and its intensity before and after the sample was measured by two ionization chambers. The harmonic rejection was achieved by the uncoated silicon plane mirror.

\begin{figure*}[t]
	\centering
	\includegraphics[width=0.9\textwidth]{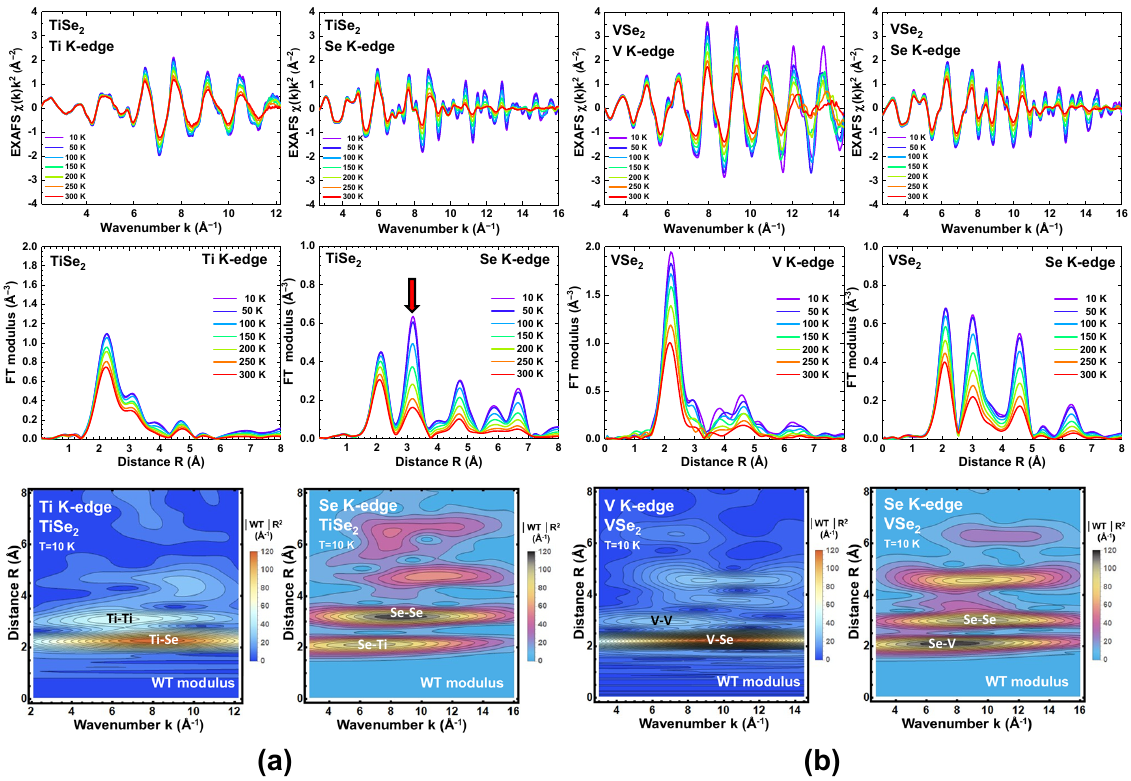}
	\caption{ Temperature-dependent EXAFS spectra $\chi(k)k^2$  (top panels), their Fourier transforms (only modulus is shown) (middle panels) and wavelet transforms at 10~K (bottom panels) for TiSe$_2$ (a) and VSe$_2$ (b). }
	\label{fig3}
\end{figure*}

Experimental Ti, V and Se K-edge EXAFS spectra were extracted following the conventional procedure \cite{Kuzmin2014} using the XAESA code \cite{XAESA}. The EXAFS spectra $\chi(k)k^2$\ and their Fourier transforms (FTs) are shown at selected temperatures  in Fig.\ \ref{fig3}. Note that the  FTs were not adjusted to compensate for the backscattering phase shift of atoms, therefore, the positions of all peaks appear shifted towards shorter distances compared to their crystallographic values.

\subsection{Reverse Monte Carlo simulations} 

The structural information encoded in the experimental EXAFS spectra was extracted using the reverse Monte Carlo (RMC) method based on an evolutionary algorithm (EA)  approach implemented in the EvAX code \cite{Timoshenko2012rmc,Timoshenko2014rmc}. The RMC/EA-EXAFS technique aims to minimize the discrepancy between  the experimental and calculated EXAFS spectra. As a result, it enables the reconstruction of the three-dimensional structural model of the material by introducing random atomic displacements within specific geometric constraints \cite{Timoshenko2012rmc,Timoshenko2014rmc}. 

In this study, the initial structural models for the RMC/EA calculations were constructed based on diffraction data \cite{Riekel1976, Hoschek1939}.
TiSe$_2$ and VSe$_2$ crystallize in the space group $P\bar{3}m1$ (164)  with the lattice parameters $a$ = $b$ = 3.540~\AA, $c$ = 6.008~\AA\  for TiSe$_2$  \cite{Riekel1976} and $a$ = $b$ = 3.355~\AA, $c$ = 6.134~\AA\  for VSe$_2$ \cite{Hoschek1939}. A supercell with a size of 6$a$$\times$6$b$$\times$4$c$, including 32 atoms, and periodic boundary
conditions  was constructed for both compounds.  At each iteration, all atoms in the supercell were displaced with the maximum allowed displacement set to 0.4~\AA. 32 atomic configurations were employed simultaneously in the EA method \cite{Timoshenko2014rmc}. 

The configuration-averaged EXAFS spectra at the  Ti/V and Se K-edges  were calculated using the ab initio self-consistent real-space multiple-scattering FEFF8.5L code \cite{Ankudinov1998,Rehr2000} taking into account multiple-scattering contributions up to the 5$^{th}$ order.  The complex energy-dependent exchange-correlation Hedin-Lundqvist potential was employed to account for inelastic effects \cite{Hedin1971}. The amplitude scaling parameter $S_0^2$ was set to 0.9 (for Ti K-edge) or 1.0 (for V and Se K-edges).

The structural model was adjusted in each  RMC/EA calculation by minimizing the difference between the Morlet wavelet transforms (WTs) \cite{Timoshenko2009wavelet} of the experimental and calculated EXAFS spectra at two absorption edges (Ti and Se for TiSe$_2$ or V and Se for VSe$_2$) simultaneously. Thus, good agreement  between the experimental and configuration-averaged EXAFS spectra was achieved in both the direct ($R$) and reciprocal ($k$) space. 

\begin{figure*}[t]
	\centering	
	\includegraphics[width=0.9\textwidth]{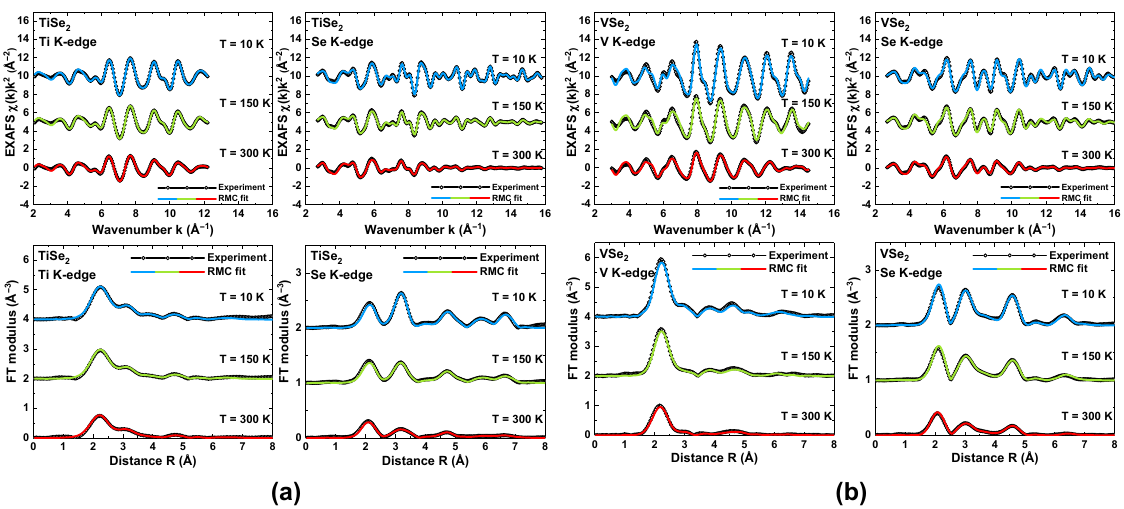}
	\caption{Results of the RMC/EA calculations for the Ti and Se K-edges in TiSe$_2$ (a) and  V and Se K-edges in VSe$_2$ (b) at selected (10, 150 and 300~K) temperatures. The EXAFS spectra $\chi(k)k^2$ are displayed within their respective fitting ranges.  The fitting $R$-space ranges were 1.5--5.9~\AA\ (Ti K-edge)   and 1.2--7.1~\AA\  (Se K-edge) 
	for TiSe$_2$  and  1.0--7.0~\AA\ (V K-edge) and 1.2--7.5~\AA\ (Se K-edge) for VSe$_2$. }
	\label{fig4}
\end{figure*}

An example of fits at the selected temperatures is shown in Fig.\ \ref{fig4}. The convergence of each RMC/EA simulation was attained after 3000 iterations.  To improve statistics, seven independent RMC/EA calculations were carried out for each experimental data set, employing distinct sequences of pseudo-random numbers. The agreement between the configuration-averaged EXAFS spectra and the experimental data at all temperatures affirms the reliability of the obtained structural models.

The atomic coordinates derived from the final RMC/EA configurations were utilized to calculate the partial radial distribution functions (RDFs) and to obtain relevant structural parameters. The mean-square relative displacement (MSRD) factors $\sigma^2$ for Ti--Se, Ti--Ti, V--Se, V--V, and Se--Se atom pairs, which account for the thermal and static disorder, were calculated using the median absolute deviation (MAD) method for the first ten coordination shells \cite{Hampel1973robust,Daszykowski2007robust}. The temperature dependencies of the obtained MSRDs were further approximated by the correlated Einstein model \cite{Sevillano1979} allowing us to determine the characteristic Einstein frequencies $\omega_E$  and the effective force constants $\kappa$, which are reported in Tables\ \ref{table1} and  \ref{table2}.

\begin{table}[t]
	\centering 
	\caption{Values of coordination numbers ($N$) and  Ti--Se, Ti--Ti, and Se--Se interatomic distances $r$  for the first ten coordination shells of Ti and Se calculated from the crystallographic structure of TiSe$_2$ \protect\cite{Riekel1976}. Interlayer distances are marked with ``*''.  The values of characteristic Einstein frequencies ($\omega_E$) and the effective force constants ($\kappa$) obtained from the RMC/EA analysis are also given for each atom pair.} 
	\label{table1}
	\begin{tabular}{llcll}
		\hline
		Atom pair & $N$  & Distance $r$ (\AA) & $\omega_E$ (THz) & $\kappa$ (N/m)\\
		\hline
		Ti$_0$--Se$_1$      &6  & 2.55  &  32.3$\pm$0.4	&  51.61$\pm$0.01  \\
		Ti$_0$--Ti$_2$      &6  & 3.54  &  31.3$\pm$0.6	&  39.02$\pm$0.02  \\
		Ti$_0$--Se$_3$      &6  & 4.37  &  23.3$\pm$0.4	&  26.92$\pm$0.01  \\
		Ti$_0$--Se$_4^*$    &6  & 4.92  &  23.3$\pm$0.6	&  26.94$\pm$0.02  \\
		Ti$_0$--Se$_5$      &12 & 5.62  &  25.6$\pm$0.9	&  32.43$\pm$0.04  \\
		Ti$_0$--Ti$_6^*$    &2  & 6.01  &  25.6$\pm$1.7 &  25.99$\pm$0.12  \\
		Ti$_0$--Se$_7^*$    &6  & 6.06  &  26.8$\pm$1.2	&  35.65$\pm$0.07  \\
		Ti$_0$--Ti$_8$      &6  & 6.13  &  32.4$\pm$0.6	&  41.69$\pm$0.01  \\
		Ti$_0$--Ti$_9^*$    &12 & 6.97  &  29.8$\pm$1.9 &  35.41$\pm$0.14  \\
		Ti$_0$--Se$_{10}^*$ &12 & 7.02  &  26.2$\pm$0.2	&  33.96$\pm$0.01  \\
		\hline
		\hline
		
		Se$_0$--Ti$_1$      &3   & 2.55  &  32.3$\pm$0.4	& 51.61$\pm$0.01  \\   
		Se$_0$--Se$_2$      &6   & 3.54  &  20.7$\pm$0.2	& 28.12$\pm$0.01  \\
		Se$_0$--Se$_3^*$    &3   & 3.58  &  28.1$\pm$1.2	& 51.95$\pm$0.10  \\
		Se$_0$--Se$_4$      &3   & 3.68  &  31.7$\pm$1.3	& 65.81$\pm$0.12  \\
		Se$_0$--Ti$_5$      &3   & 4.37  &  23.3$\pm$0.4	& 26.92$\pm$0.01  \\
		Se$_0$--Ti$_6^*$    &3   & 4.92  &  23.3$\pm$0.6	& 26.94$\pm$0.02  \\
		Se$_0$--Se$_7^*$    &3   & 5.04  &  22.4$\pm$0.5	& 32.89$\pm$0.02  \\
		Se$_0$--Se$_8$      &3   & 5.11  &  24.7$\pm$0.6	& 39.96$\pm$0.03  \\
		Se$_0$--Ti$_9$      &6   & 5.62  &  25.6$\pm$0.9	& 32.43$\pm$0.04  \\
		Se$_0$--Se$_{10}^*$ &2   & 6.01  &  24.6$\pm$0.8	& 39.53$\pm$0.04  \\
		\hline
	\end{tabular}
\end{table}

\begin{table}[t]
	\centering 
	\caption{Values of coordination numbers ($N$) and V--Se, V--V, and Se--Se interatomic distances $r$  for the first ten coordination shells of V and Se calculated from the crystallographic structure of VSe$_2$ \protect\cite{Hoschek1939}. Interlayer distances are marked with ``*''. Coordination shells composed of both intralayer and interlayer atomic pairs are marked with ``**''. The values of characteristic Einstein frequencies ($\omega_E$) and the effective force constants ($\kappa$) obtained from the RMC/EA analysis are also reported for each atom pair.} 
	\label{table2}
	\begin{tabular}{llcll}
		\hline
		Atom pair         &$N$  & Distance $r$  (\AA) & $\omega_E$ (THz) & $\kappa$ (N/m)\\
		\hline
		V$_0$--Se$_1$     &6 & 2.47  &  31.1$\pm$0.8	&  49.69$\pm$0.04  \\
		V$_0$--V$_2$      &6 & 3.36  &  16.6$\pm$0.4	&  11.67$\pm$0.01  \\
		V$_0$--Se$_3$     &6 & 4.17  &  21.7$\pm$0.4	&  24.25$\pm$0.01  \\
		V$_0$--Se$_4^*$   &6 & 4.99  &  24.9$\pm$1.0	&  31.84$\pm$0.06  \\
		V$_0$--Se$_5$     &12& 5.35  &  21.3$\pm$0.6	&  23.23$\pm$0.02  \\
		V$_0$--V$_6$      &6 & 5.81  &  29.5$\pm$2.3 	&  36.81$\pm$0.22  \\
		V$_0$--Se$_7^*$   &6 & 6.01  &  25.4$\pm$1.6	&  33.13$\pm$0.13  \\
		V$_0$--V$_8^*$    &2 & 6.13  &  28.4$\pm$1.3	&  34.03$\pm$0.07  \\
		V$_0$--V$_9$      &6 & 6.71  &  20.1$\pm$1.2 	&  17.16$\pm$0.06  \\
		V$_0$--Se$_{10}^*$&12& 6.89  &  26.2$\pm$1.1	&  35.22$\pm$0.06  \\
		\hline
		\hline
		
		Se$_0$--V$_1$          &3  & 2.47  &  31.1$\pm$0.8	& 49.69$\pm$0.04  \\
		Se$_0$--Se$_2$         &6  & 3.36  &  24.5$\pm$0.4	& 39.32$\pm$0.01  \\
		Se$_0$--Se$_{3/4}^{**}$&3+3& 3.63  &  27.1$\pm$1.4	& 48.18$\pm$0.12  \\
		Se$_0$--V$_5$          &3  & 4.17  &  21.7$\pm$0.4	& 24.25$\pm$0.01  \\
		Se$_0$--Se$_{6/7}^{**}$&3+3& 4.94  &  25.1$\pm$0.6	& 41.19$\pm$0.02  \\
		Se$_0$--V$_8$          &3  & 4.99  &  24.9$\pm$1.0	& 31.84$\pm$0.06  \\
		Se$_0$--V$_9$          &6  & 5.35  &  21.3$\pm$0.6	& 23.23$\pm$0.02  \\
		Se$_0$--Se$_{10}$      &6  & 5.81  &  23.5$\pm$0.4	& 36.16$\pm$0.01  \\
		%		Se$_0$--Se$_{11}$      &12 & 5.97  &  26.1$\pm$0.5	& 44.78$\pm$0.01  \\
		\hline
	\end{tabular}
\end{table}

\section{Results and discussion}

The crystallographic structure of TiSe$_2$ (VSe$_2$)  (Fig.\ \ref{fig1}) consists of a titanium (vanadium) atomic layer sandwiched between two selenium atomic layers \cite{Riekel1976, Hoschek1939}. These layers are weakly bonded to each other along the $c$-axis through vdW forces. The interlayer gap between Se atoms is about 3.12~\AA\ in TiSe$_2$  and 3.16~\AA\ in VSe$_2$. Ti(V) atoms are covalently bonded to six selenium atoms, forming regular octahedra. The values of Ti--Se, Ti--Ti, and Se--Se interatomic distances for the first ten coordination shells in TiSe$_2$ are reported in Table\ \ref{table1}, while the values of  V--Se, V--V, and Se--Se interatomic distances in VSe$_2$ are summarized in Table\ \ref{table2}. 

Despite the similarity between the two crystallographic structures, a notable difference can be observed as well. In the case of VSe$_2$, the interatomic distances between selenium atoms along the $c$-axis direction within the same layer and across adjacent layers coincide ($r_{{\rm Se}_0-{\rm Se}_{3/4}^{**}}$ = 3.63~\AA), adding complexity to the analysis. In contrast, in TiSe$_2$, the Se$_3^*$ atoms in the adjacent layer are approximately 0.1~\AA\ closer to the absorbing Se$_0$ than the Se$_4$ atoms within the same layer, so that the two interatomic distances are different 
$r_{{\rm Se}_0-{\rm Se}_3^*}$ = 3.58~\AA\ and $r_{{\rm Se}_0-{\rm Se}_4}$ = 3.68~\AA.

The Ti K-edge EXAFS spectra of TiSe$_2$ and their FTs are dominated by a contribution from the first two coordination shells at all studied temperatures in the range of 10-300~K (Fig.\ \ref{fig3}(a)). These coordination shells consist of six selenium  and  six titanium  atoms, respectively. The complex pattern of WTs observed at longer distances is attributed to outer shells and multiple-scattering effects. Note that heavier Se atoms with an atomic mass of 78.971~amu produce a greater impact on the EXAFS spectra at larger $k$ values, whereas lighter Ti elements with an atomic mass of 47.867~amu contribute at lower $k$-values. 

The Se K-edge EXAFS spectra of TiSe$_2$  contain contributions from scattering paths that extend beyond $\sim$7~\AA\ as evident in the FTs and WTs in Fig.\ \ref{fig3}(a). Their analysis can yield novel insights into the interactions between layers \cite{Pudza2023}.  Indeed, the second peak at about 3.2~\AA\  in FTs (indicated with a red arrow in Fig.\ \ref{fig3}(a)) contains contributions from intralayer ($r_{{\rm Se}_0-{\rm Se}_2}$ = 3.54~\AA, $r_{{\rm Se}_0-{\rm Se}_4}$ = 3.68~\AA) and interlayer ($r_{{\rm Se}_0-{\rm Se}_3^*}$ = 3.58~\AA)  selenium atoms, and its amplitude  becomes significantly suppressed upon sample heating. Note that the amplitude of other peaks up to $\sim$7~\AA\ is also reduced upon increasing thermal disorder but the peaks remain distinguishable even at 300~K.

While the reliable analysis of distant coordination shells is challenging within the conventional EXAFS methodology \cite{Kuzmin2014}, it can be performed using the RMC/EA method \cite{Pudza2023, Timoshenko2014rmc, Kuzmin2019}. Examples of the RMC/EA simulations of the Ti and Se K-edge EXAFS spectra for TiSe$_2$ at three selected temperatures (10, 150 and 300~K) are shown in Fig.\ \ref{fig4} in $k$- and $R$-space. Good agreement between experimental and calculated EXAFS spectra is observed, enabling a comprehensive analysis of thermal disorder effects within the first ten coordination shells up to about 6~\AA.  

\begin{figure*}[t]
	\centering
	\includegraphics[width=0.9\textwidth]{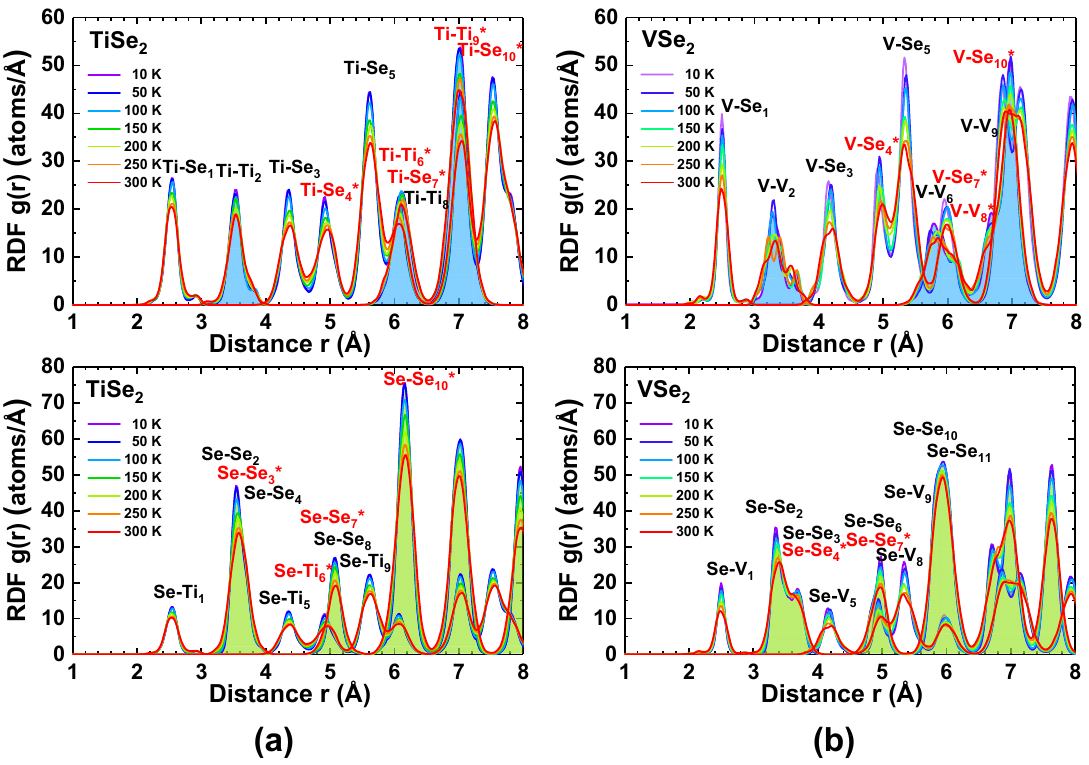}
	\caption{ Partial radial distribution functions (RDFs) around Ti and Se atoms in TiSe$_2$ (a) and V and Se atoms in VSe$_2$ (b) as a function of temperature. Open and filled peaks correspond to different partial RDFs. }
	\label{fig5}
\end{figure*}

\begin{figure*}[t]
	\centering
	\includegraphics[width=0.6\textwidth]{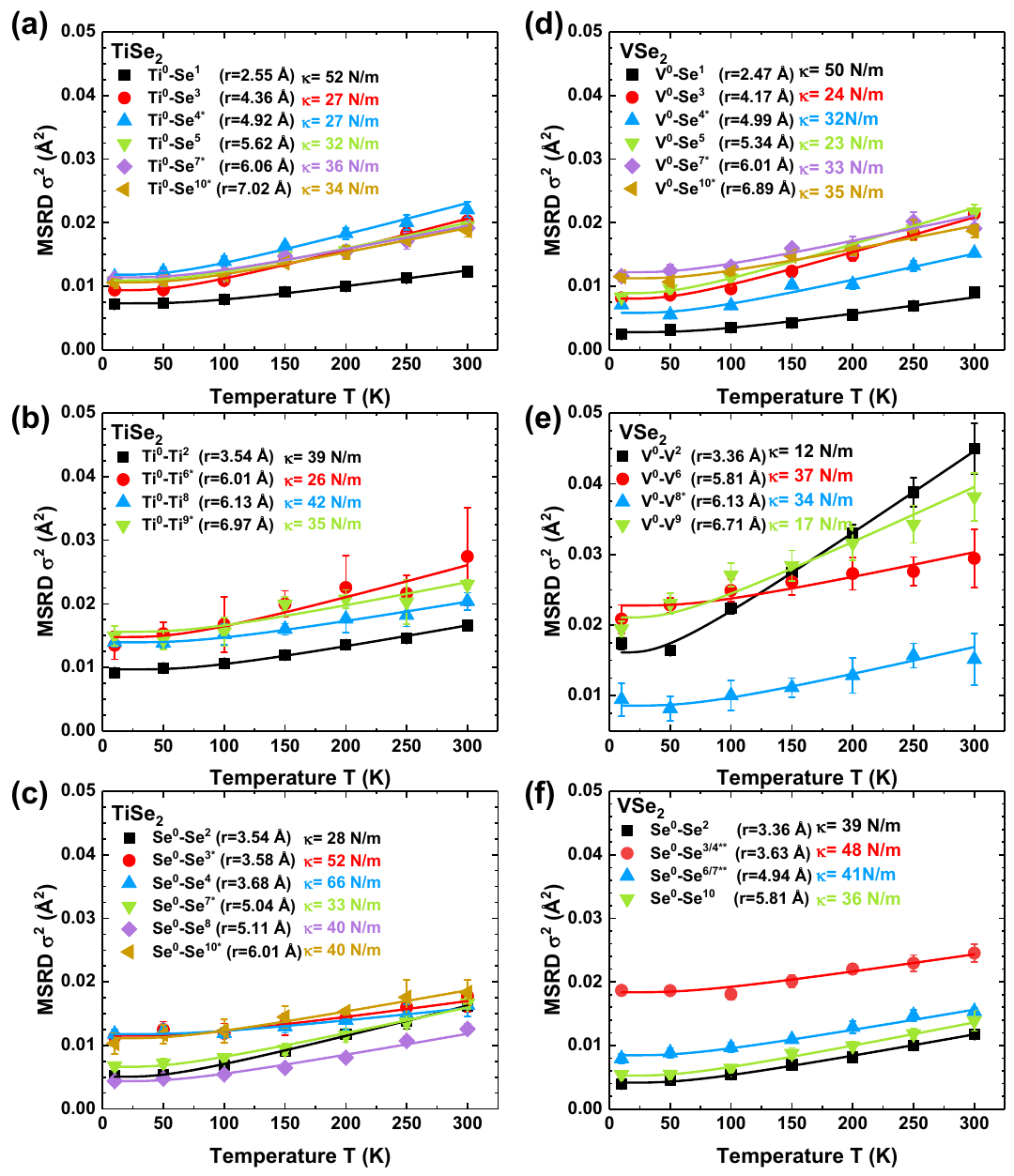}
	\caption{ Temperature dependence of the mean-square relative displacement (MSRD) factors $\sigma^2$  for Ti(V)--Se (a,d), Ti--Ti(V--V) (b,e), and Se--Se (c,f) atom pairs in TiSe$_2$ and VSe$_2$. Solid lines correspond to the fits using the correlated Einstein model \protect\cite{Sevillano1979}. Calculated values of the effective force constants $\kappa$ are also given.  }
	\label{fig6}
\end{figure*}

The coordinates of atoms derived from the RMC/EA simulations were employed to calculate the partial partial RDFs $g(r)$  (Fig.\ \ref{fig5})  and the MSRDs $\sigma^2$ (Fig.\ \ref{fig6}) for Ti--Ti, Ti--Se(Se--Ti), and Se--Se atom pairs at each  temperature. The temperature dependencies of the obtained MSRDs  $\sigma^2(T)$ in the range of 10-300~K were further  approximated using the correlated Einstein model \cite{Sevillano1979}. As a result, the characteristic Einstein frequencies $\omega_E$  and the effective force constants $\kappa$ were obtained for all coordination shells and are reported in Table\ \ref{table1}.

The  effective force constants $\kappa$ for the first two coordination shells of titanium (Ti$_0$--Se$_1$ and Ti$_0$--Ti$_2$) are about 52~N/m and 39~N/m, respectively, suggesting strong interaction between the nearest atoms. The interactions between absorbing Ti$_0$ and the next Se atoms located in the same (Se$_3$) or in the adjacent layer (Se$_4^*$) are close, as indicated by similar values of $\kappa \approx 27$~N/m. 
At the same time, the intralayer Ti--Ti interactions (Ti$_0$--Ti$_2$, Ti$_0$--Ti$_8$) are slightly stronger ($\kappa \approx 39$-42~N/m) than the interlayer Ti--Ti interactions  (Ti$_0$--Ti$_6^*$, Ti$_0$--Ti$_9^*$) with $\kappa \approx 26$-35~N/m.
Among all Se--Se atom pairs, the interaction between the nearest selenium atoms is the weakest ($\kappa_{{\rm Se}_0-{\rm Se}_2} \approx 28$~N/m), while the interaction between two selenium atoms located at opposite sides of the same layer is more than twice stronger ($\kappa_{{\rm Se}_0-{\rm Se}_4} \approx 66$~N/m). At the same time, the interaction between selenium atoms in adjacent layers across the vdW gap is rather strong with $\kappa_{{\rm Se}_0-{\rm Se}_3^*} \approx 52$~N/m. 
To conclude, the temperature dependencies of the MSRD factors in TiSe$_2$ do not show any unusual behaviour.

In spite of the similarities in the crystallographic structures of TiSe$_2$ and VSe$_2$, the EXAFS spectra $\chi(k)k^2$ of VSe$_2$ have stronger amplitude and, as a result, more intense peaks in FTs (Fig.\ \ref{fig3}).   
The V K-edge EXAFS spectra of VSe$_2$, their FTs and WTs (Fig.\ \ref{fig3}(b)) are dominated by a contribution from the first coordination shell composed of six selenium atoms. Contributions from distant coordination shells are also present up to 7~\AA\ but they are relatively weaker. This fact contrasts with the Se K-edge EXAFS spectra where strong contributions from distant shells are observed in FTs up to 5~\AA\ but appreciable signals are visible even at longer distances. 

The results of the RMC/EA fits for VSe$_2$ are in good agreement with the experimental data at both V and Se K-edges (Fig.\ \ref{fig4}(b)). The partial RDFs for V--V, V--Se(Se--V), and Se--Se atom pairs are shown in Fig.\ \ref{fig5}(b), whereas the characteristic Einstein frequencies $\omega_E$  and the effective force constants $\kappa$ for all coordination shells  are reported in Table\ \ref{table2}. 

The RDFs obtained for both TMDs (Fig.\ \ref{fig5})  exhibit an expected temperature dependence:  they become broadened at higher temperatures. At the same time, no peak splitting is observed for the nearest shells at low temperatures in the CDW state. This suggests that the expected lattice distortions due to atom displacements \cite{DiSalvo1976,Guo2021} are relatively small compared to thermal disorder effects. 
Nevertheless, upon comparing the partial  RDFs for TiSe$_2$ and VSe$_2$, a notable difference in the temperature dependence between Ti-Ti and V-V RDFs is observed within the range of the second coordination shell. Upon increasing temperature, the V$_0$--V$_2$ RDF peak at 3.36~\AA\ in  VSe$_2$ becomes significantly more broadened compared to the Ti$_0$--Ti$_2$ RDF peak at 3.54~\AA\ in TiSe$_2$. This difference is well reflected by the temperature dependence of respective MSRDs in Fig.\ \ref{fig6}(b) and (e).  Indeed, the MSRD for the V$_0$--V$_2$ atom pair exhibits a rapid increase above 50~K, displaying anomalous behavior compared to MSRDs for other atom pairs.   
Note that the value of the force constant  $\kappa_{{\rm V}_0-{\rm V}_2} \approx 12$~N/m  is the smallest one (Table\ \ref{table2}).
    
There is also a prominent difference in the splitting of the peak around 3.6~\AA\ in the Se--Se RDFs between the two TMDs. A single peak at 3.55~\AA\ is present in TiSe$_2$, whereas a double peak at about 3.35~\AA\ and 3.68~\AA\ exists in VSe$_2$.  This difference is due that the Se$_0$--Se$_2$ distance in VSe$_2$ is significantly shorter by 0.18~\AA\ than in TiSe$_2$  (see Tables\ \ref{table1} and \ref{table2}).

\section{Conclusions}

Highly crystalline bulk 1T-TiSe$_2$ and 1T-VSe$_2$ were synthesized using the chemical vapor transport technique, and their local atomic structure and lattice dynamics were  studied by  temperature-dependent (10-300~K) X-ray absorption spectroscopy at the Ti, V, and Se K-edges. 

The extended X-ray absorption fine structure (EXAFS) spectra were analysed using the reverse Monte Carlo (RMC) simulations \cite{Timoshenko2012rmc,Timoshenko2014rmc} by a simultaneous fitting of metal (Ti/V) and selenium K-edge  spectra  (Fig.\ \ref{fig4}). Such an approach allowed us to reconstruct the local environment in both selenides in terms of the partial radial distribution functions  (Fig.\ \ref{fig5}) and to gain reliable information on disorder effects in the nearest and distant coordination shells. The mean squared relative displacements $\sigma^2$ for atom pairs located in the first ten coordination shells were determined, and their temperature dependencies were approximated using the correlated Einstein model.  As a result, the characteristic Einstein frequencies $\omega_E$ and the effective force constants $\kappa$ were obtained (Tables\ \ref{table1} and \ref{table2}) for all atomic pairs and used in the assessment of the strengths of intralayer and interlayer interactions. 

The partial RDFs obtained for both TMDs (Fig.\ \ref{fig5}) demonstrate the expected temperature dependence and do not show any evidence of the peak splitting for the nearest shells at low temperatures due to the CDW state. This indicates that CDW-induced atom displacements \cite{DiSalvo1976,Guo2021} are  relatively small compared to thermal disorder effects. 
Nevertheless, a comparison of the RDFs for Ti--Ti and V--V atom pairs located within the same layer shows their different temperature variations.  In particular, the thermal disorder significantly affects the V$_0$--V$_2$ atom pair as can be seen from the temperature dependence of its MSRD  (Fig.\ \ref{fig6}) and small value of the force constant  $\kappa_{{\rm V}_0-{\rm V}_2} \approx 12$~N/m  (Table\ \ref{table2}).

Thus, the RMC/EA-EXAFS technique shows great potential for studying the local atomic structure and disorder effects in TMDs and other layered materials.

\section*{CRediT authorship contribution statement}

\textbf{Inga Pudza}: Investigation, Visualization, Writing -- original draft, Writing -- review \& editing.
\textbf{Boris Polyakov}:  Investigation, Writing -- original draft, Writing -- review \& editing.
\textbf{Kaspars Pudzs}: Investigation.
\textbf{Edmund Welter}: Resources.
\textbf{Alexei Kuzmin}: Conceptualization, Investigation, Methodology, Writing -- original draft, Writing -- review \& editing.

\section*{Declaration of Competing Interest}

The authors declare that they have no known competing financial interests or personal relationships that could have appeared to influence the work reported in this paper.

\section*{Acknowledgements}

B.P. and A.K. thank the support of the Latvian Council of Science project No. LZP-2020/1-0261.
The experiment at the PETRA III synchrotron was performed within proposal No. I-20210625 EC.
The synchrotron experiment has been supported by the project CALIPSOplus under the Grant Agreement 730872 from the EU Framework Programme for Research and Innovation HORIZON 2020.
Institute of Solid State Physics, University of Latvia as the Center of Excellence has received funding from the European Union's Horizon 2020 Framework Programme H2020-WIDESPREAD-01-2016-2017-TeamingPhase2 under grant agreement No. 739508, project CAMART2.

\section*{Data availability statement}

Data will be made available on request.

\newpage

% \bibliography{2DCDW}

\end{document}